\documentclass[sigconf]{acmart}
\AtBeginDocument{%
  }

\copyrightyear{2025}
\acmYear{2025}
\setcopyright{rightsretained}
\acmConference[UIST Adjunct '25]{The 38th Annual ACM Symposium on User Interface Software and Technology}{September 28-October 1, 2025}{Busan, Republic of Korea}
\acmBooktitle{The 38th Annual ACM Symposium on User Interface Software and Technology (UIST Adjunct '25), September 28-October 1, 2025, Busan, Republic of Korea}\acmDOI{10.1145/3746058.3758362}
\acmISBN{979-8-4007-2036-9/2025/09}

\usepackage{xspace}
\usepackage{listings}
\newcommand{\SystemName}{\textsc{AdjustAR}\xspace}

\begin{document}

\title{AdjustAR: AI-Driven In-Situ Adjustment of Site-Specific Augmented Reality Content}

\author{Nels Numan}
\affiliation{%
  \institution{University College London}
  \city{London}
  \country{United Kingdom}
  }
\email{nels.numan@ucl.ac.uk}

\author{Jessica Van Brummelen}
\affiliation{%
  \institution{Niantic Spatial, Inc.}
  \city{London}
  \country{United Kingdom}
  }
\email{jess@nianticspatial.com}

\author{Ziwen Lu}
\affiliation{%
  \institution{University College London}
  \city{London}
  \country{United Kingdom}
  }
\email{ziwen.lu@ucl.ac.uk}

\author{Anthony Steed}
\affiliation{%
  \institution{University College London}
  \city{London}
  \country{United Kingdom}
  }
\email{a.steed@ucl.ac.uk}

\renewcommand{\shortauthors}{Numan et al.}

\begin{abstract}
Site-specific outdoor AR experiences are typically authored using static 3D models, but are deployed in physical environments that change over time. As a result, virtual content may become misaligned with its intended real-world referents, degrading user experience and compromising contextual interpretation. We present \SystemName, a system that supports in-situ correction of AR content in dynamic environments using multimodal large language models (MLLMs). Given a composite image comprising the originally authored view and the current live user view from the same perspective, an MLLM detects contextual misalignments and proposes revised 2D placements for affected AR elements. These corrections are backprojected into 3D space to update the scene at runtime. By leveraging MLLMs for visual-semantic reasoning, this approach enables automated runtime corrections to maintain alignment with the authored intent as real-world target environments evolve.
\end{abstract}


\begin{teaserfigure}
\centering
  \includegraphics[width=\textwidth]{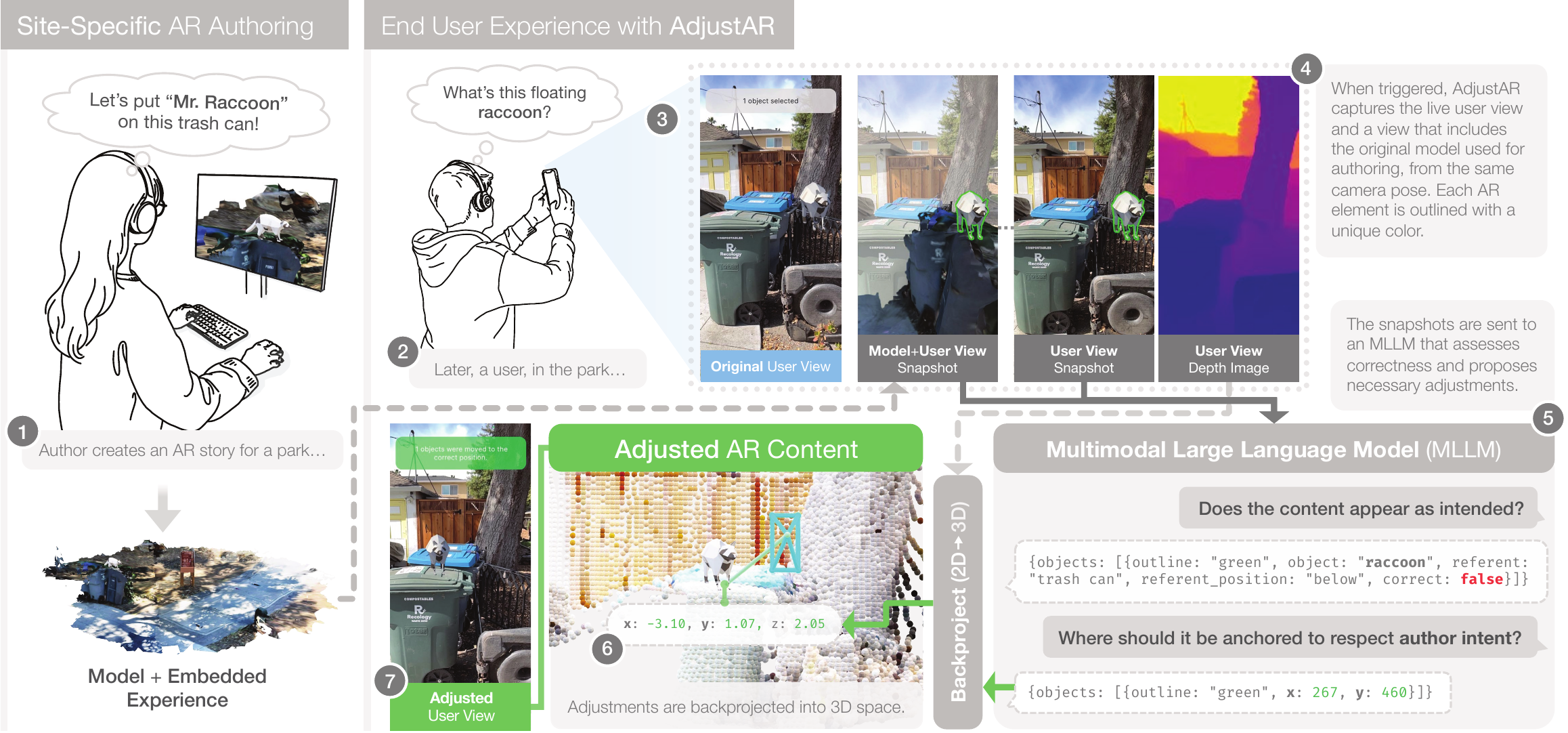}
  \caption{\SystemName corrects misaligned AR content at runtime: (1) authors place content relative to a georeferenced 3D model of the target site; (2–3) users localize and view the scene \textit{in-situ}, where misalignments may occur due to environmental changes; (4) the system composites live and authored views; (5) an MLLM detects misalignments and infers corrected 2D anchors; (6) corrections are backprojected into 3D and updated in the scene; (7) the user's AR view is updated.}
  \label{fig:teaser}
\end{teaserfigure}

\maketitle

\section{Introduction}

Model-based authoring is the dominant approach for designing outdoor site-specific AR experiences~\cite{numanCoCreatAREnhancingAuthoring2025}. Remote authoring tools~\cite{lightship-ardk-niantic,googleGeospatialCreator} enable anchoring virtual content relative to georeferenced 3D models via visual positioning systems (VPS). However, these models are static. Changes in the physical environment can cause misalignments between virtual content and its real-world referents, degrading user experience and requiring repeated site visits~\cite{numanCoCreatAREnhancingAuthoring2025}.

While prior systems have explored runtime adaptation through semantic constraints or layout optimization~\cite{lindlbauerContextAwareOnlineAdaptation2019,evangelistabeloAUITAdaptiveUser2022,chengSemanticAdaptOptimizationbasedAdaptation2021,liSituationAdaptContextualUI2024}, these efforts largely focused on user interfaces. \textit{ScalAR}~\cite{qianScalARAuthoringSemantically2022} applied this concept to AR authoring, introducing semantic-driven virtual proxies that supported model-based authoring. \citet{liInteractiveAugmentedReality2022} presented a system that extracted semantic structure from point clouds to retarget interactive narratives through story graph sampling. However, both approaches remained limited to indoor environments with narrow semantic vocabularies and relied on offline processing.

Recent advances in large language models (LLMs) and their multimodal variants (MLLMs) enable spatially grounded visual-semantic reasoning~\cite{chenSpatialVLMEndowingVisionlanguage2024,yangThinkingSpaceHow2025}. Recent work has begun to explore LLMs for mixed-reality authoring~\cite{delatorreLLMRRealtimePrompting2024,giunchiDreamCodeVRDemocratizingBehavior2024}, with \textit{ImaginateAR}~\cite{leeImaginateARAIAssistedInSitu2025} extending this concept to outdoor contexts through complex scene understanding and asset generation pipelines. While these systems expand the expressive potential of authoring, they have not focused on maintaining spatial consistency when the target environment changes.

To address this gap, we present \SystemName, a system that combines model-based authoring with MLLM-guided visual-semantic correction. Authors follow standard creation workflows, while at runtime, \SystemName compares rendered and live views to identify and correct misalignments that disrupt the intended experience. Rather than supporting re-authoring~\cite{numanCoCreatAREnhancingAuthoring2025}, \SystemName aims to adaptively preserve the original design, treating the original authored scene as the canonical expression of author intent.

\section{\SystemName System}

\SystemName extends standard site-specific AR pipelines with a runtime correction mechanism driven by MLLMs (Fig.~\ref{fig:teaser}). Authors create experiences using the Niantic SDK for Unity~\cite{lightship-ardk-niantic}, placing AR elements relative to site-specific 3D models. At runtime, users localize via VPS to align the scene with their view. When the physical target environment is unchanged, AR content appears correctly. However, when referents have moved or disappeared, visual misalignments may occur. To handle such cases, \SystemName introduces a visual-semantic correction process that aims to address misalignments and restore author intent.

\paragraph{\textbf{Visual Comparison for In-Situ Adjustment}}

\SystemName performs a combined evaluation and adjustment process that can be triggered manually or periodically. When triggered, the system captures two snapshots from the current camera pose: the live AR view and a synthetic rendering of the authored scene, generated using the model that was originally used for authoring. Virtual elements are uniquely color-coded for content-agnostic object references.

The system concurrently caches the camera intrinsics, extrinsics, and depth map for each snapshot. These are later used to compute corrected 3D anchors once MLLM feedback is available, ensuring that repositioning aligns with the visual context at trigger time, even if the user has since moved.

The two images are passed to an MLLM (\textit{Gemini 2.5 Flash}~\cite{googleGeminiFlash252025}) to assess whether each element appears aligned with its physical referent. For misaligned elements, another MLLM (\textit{Gemini 2.5 Pro}~\cite{googleGeminiPro252025}) is prompted to provide corrected 2D anchor points in image space, optionally including a 3D vertical offset. Both responses are returned in a JSON format based on a pre-defined schema. If the MLLM determines that a physical referent is no longer present or is fully occluded, \SystemName displays a rendered snapshot of the original authored experience to provide the user with an indication of the intended experience. Prompts are detailed in Appendix~\ref{sec:prompt}.

\paragraph{\textbf{3D Repositioning via Backprojection}}

Corrected 2D anchors are backprojected into 3D world coordinates using the depth map and camera information cached at trigger time. The resulting 3D point becomes the new anchor, located at the bottom center of the element's bounding box. Optional vertical 3D offsets are applied to support elevated placements (e.g., hovering arrows). Finally, the AR scene is updated to reflect the adjusted anchor positions and is displayed to the user.

\section{Future Work}
Future work on \SystemName should address system performance, authoring support, and empirical validation, including both quantitative evaluation and user studies in diverse deployment contexts.

Performance improvements may include reducing the correction pipeline latency (currently $\sim$10-20 seconds) and improving accuracy, potentially through prompt optimization and additional input modalities such as depth, multi-view, or mesh data~\cite{yehSeeingAnotherPerspective2025}. Advances in MLLMs, especially in spatial reasoning and 3D grounding, are likely to support these improvements~\cite{chenSpatialVLMEndowingVisionlanguage2024,yangThinkingSpaceHow2025}.

The current correction mechanism operates on static image pairs where referents and AR elements are visible within the same frame. Future work could incorporate spatiotemporal observations to handle occlusion or out-of-frame references. When referents are missing, virtual proxies may help preserve semantic continuity.

Placement decisions are currently based on a bottom-center heuristic, where the MLLM anchors objects relative to their base with an optional vertical offset. Future work could explore other anchoring strategies referencing surfaces, edges, or other geometric features~\cite{nuernbergerSnapToRealityAligningAugmented2016,qianScalARAuthoringSemantically2022}. While prior work has examined semantically meaningful placements in narratives~\cite{liInteractiveAugmentedReality2022,liLocationAwareAdaptationAugmented2023}, how closely AR content should align with target environments requires further exploration. For example, for a site-specific AR story, a character placed by a specific tree might appear near a similar one if the original is absent, whereas training applications may require exact replication.

Finally, supporting author-defined semantic constraints (e.g., ``must be visible from entrance'') could enable more precise intent specification and guide adaptation to contextual changes such as crowdedness, seasonal change, or other situational changes~\cite{liSituationAdaptContextualUI2024}. MLLMs offer a mechanism for interpreting such constraints flexibly, enabling adaptive behavior from sparse multimodal input.

\begin{acks}
This work was partially supported by the European Union's Horizon 2020 Research and Innovation program as part of project RISE under grant agreement No. 739578. We thank Simon Julier, Gabriel Brostow, and Niladri Dutt for valuable research discussions, and Dat Chu for assistance with testing logistics. The raccoon model used in this work was sourced from Google Poly.
\end{acks}

\bibliographystyle{ACM-Reference-Format}
\bibliography{references_adjustAR}

\appendix

\section{System Prompts}\label{sec:prompt}

\subsection{Initial Check for Correct Alignment (\textit{Gemini Flash})}
\begin{lstlisting}
You are a visual analysis AI agent. You are given a side-by-side image showing two versions of the same augmented reality (AR) scene:
-   Left image: AR as captured in real-world use.
-   Right image: AR as authored and intended to appear.

For each outlined object, assess whether its placement in the left image matches its placement in the right image, relative to its physical referent. A physical referent is the real-world object, surface, or spatial location to which the AR content is anchored or aligned. It provides the spatial or semantic basis for interpreting the AR element in the physical environment and is often the nearest visible surface or object.

For each object, indicate:
* The name of the physical referent.
* The position of the referent relative to the outlined object in the right image (from the camera's perspective).
* Whether the placement in the left image is correct.
* Whether the physical referent is visible in the left image.
\end{lstlisting}
\clearpage
\subsection{Request for Adjusted Anchor Points (\textit{Gemini Pro})}
\begin{lstlisting}
You are a visual analysis AI agent. You are given a side-by-side image showing two versions of the same augmented reality (AR) scene:
- Left image: AR as captured in real-world use.
- Right image: AR as authored and intended to appear.

Your task is to determine whether each AR element (e.g., arrows, labels, icons) in the left image is correctly aligned with the same physical referent as in the right image.

A physical referent is the real-world object, surface, or spatial location to which AR content is anchored or aligned. It provides the spatial or semantic basis for interpreting the AR element in the physical environment and is typically the nearest visible surface or object.

If an AR element in the left image is misaligned, your task is to provide a corrected anchor position directly on the physical referent in the left image. If needed, also specify a vertical Y offset (in centimeters) indicating how far above or below this point the AR element should appear in 3D space. Your task is as follows:

1. Identify misalignments
Examine all AR elements in the left image. Each element is outlined in a unique color: "green", "blue", "magenta", "red", "orange", "yellow", or "cyan". For each element, assess whether it is anchored to the same physical referent as in the right image.

2. Correct misaligned elements
For each element that is not aligned with the correct referent:

- Specify a corrected anchor point in the left image.
- This point must lie directly on the same component of the physical referent, not near it or floating above it.
- Prioritize spatial accuracy. The position must align with the physical referent even if that referent has moved or changed appearance.
- Emphasize local visual consistency: prefer alignment with the object or surface the AR element was originally intended to refer to, rather than matching unrelated global features such as sky, shadows, or pavement.
- If either image is visually degraded or ambiguous, make the best possible contextual inference about the referent's location based on visible cues.
- Provide coordinates as follows:
* X and Y: Normalized to the left image dimensions, ranging from 0 to 1000. Origin (0, 0) is top-left; (1000, 1000) is bottom-right.
* Y offset (in centimeters): A vertical offset in 3D space from the anchor point. Positive values are upward, negative values are downward. If the element is placed directly on the referent or the vertical offset is unknown, use `0`.

3. Handle missing physical referents
If an AR element in the right image is meant to point to a physical referent that is no longer present or fully occluded in the left image, mark the element as "Missing".

4. Skip correct elements
If the AR element in the left image correctly points to or aligns with the same referent as in the right image, mark it as "Correct". Do not suggest changes.
\end{lstlisting}
\end{document}